\documentclass[conference]{IEEEtran}
\usepackage{cite}

\usepackage[cmex10]{amsmath}

\usepackage{times,amsmath,epsfig}
\usepackage{epstopdf}
\usepackage{psfrag}
\usepackage{subfigure}
\usepackage[absolute]{textpos}
\newcommand{\copyrightstatement}{
    \begin{textblock}{15}(0.5,0.3)    
         \noindent
         \centering
         \textblockcolour{white}
         \footnotesize
         \copyright 2016 IEEE. Personal use of this material is permitted. Permission from IEEE must be obtained for all other uses, in any current or future media, including reprinting/republishing this material for advertising or promotional purposes, creating new collective works, for resale or redistribution to servers or lists, or reuse of any copyrighted component of this work in other works
    \end{textblock}
}

\begin{document}
%
\title{Joint Optimization of Rate, Distortion, and Decoding Energy for HEVC Intraframe Coding}

\copyrightstatement

\author{\IEEEauthorblockN{Christian Herglotz and Andr\'e Kaup}
\IEEEauthorblockA{Multimedia Communications and Signal Processing\\
Friedrich-Alexander University Erlangen-N\"urnberg,
Cauerstr. 7, 91058 Erlangen, Germany\\ Email: \{ \ christian.herglotz, andre.kaup\ \}\ @~FAU.de}}

\maketitle

\begin{abstract}
This paper presents a novel algorithm that aims at minimizing the required decoding energy by exploiting a general energy model for HEVC-decoder solutions. 
We incorporate the energy model into the HEVC encoder such that it is capable of constructing a bit stream whose decoding process consumes less energy than the decoding process of a conventional bit stream. To achieve this, we propose to extend the traditional Rate-Distortion-Optimization scheme to a Decoding-Energy-Rate-Distortion approach. To obtain fast encoding decisions in the optimization process, we derive a fixed relation between the quantization parameter and the Lagrange multiplier for energy optimization. Our experiments show that this concept is applicable for intraframe-coded videos and that for local playback as well as online streaming scenarios, up to $\mathbf{15\%}$ of the decoding energy can be saved at the expense of a bitrate increase of approximately the same magnitude. 
\end{abstract}

\begin{IEEEkeywords}
HEVC, energy, decoder, model, rate-distortion-optimization.
\end{IEEEkeywords}

%
\IEEEpeerreviewmaketitle

\section{Introduction}
Nowadays, portable devices such as smartphones or tablet PCs have become an indispensable gadget for millions of users all over the world. A major drawback of these devices is that the battery capacity in conjunction with the power consumption of the system limits the operating time of the device. Hence, research aiming at reducing the power consumption of a process is a worthwhile task. 

This paper addresses the power consumption of the video decoding process. Analyzing the power consumption of a portable device it was shown that during video playback, the decoding process consumes a major part of the battery power \cite{Carroll10}. To reduce the power consumption, great effort is pursued to optimize decoder implementations, not only on the software side \cite{FFmpeg}, but also on the hardware side \cite{Tikekar14, Kalali12, Adibelli11}. Using these techniques, energy can be saved in the range of several orders of magnitude. In this paper we proceed in a different direction 
and tackle the following question: If we have given a fixed (and maybe optimized) decoder solution, is it possible to further reduce the decoding energy using knowledge about its energetic properties? In this paper, we show that for intraframe coded videos this is indeed possible. 

To address this problem, we propose to construct bit streams that consume less energy during the decoding process than conventional bit streams. To this end, we provide information about the energetic properties of the decoder to the encoder such that it can exploit this knowledge to decide for energy saving coding modes. This paper investigates how much decoding energy can be saved when such knowledge is available. 

As in this context the decision criterion of the encoder is changed, the resulting bit streams may have a lower rate-distortion performance than before. That means that we trade decoding energy against bitrate. At first glance this may seem counter-intuitive as it is usually assumed that a larger bit stream requires more energy during decoding. In contrast we argue that depending on the encoding decision, the decoding process may show a different complexity. E.g., we can encode a so-called transform-skip flag. If the flag is false, an inverse, computationally complex transform is performed requiring a certain amount of processing energy. If the flag is true, the transform is skipped and thus no additional processing energy is required. We can see that, although both decisions are expressed by only one bit, the required processing energy on the decoder side may differ significantly. 


There were similar works in the past where power or processing complexity, which can be interpreted as an approximation to processing power, was investigated. E.g., Zhihai He et al. \cite{He05} incorporated the encoding power into the Rate-Distortion calculation to save energy on the encoder side. Using the encoder's processing complexity, this idea has been deeply investigated, e.g., by Xiang Li et al. \cite{Li11}. In another work, Yuwen He et al. \cite{He13} proposed a simple decoding complexity model for the modules deblocking filter and motion compensation and showed that operating times can be extended using power-aware optimization criteria. A more complex model for the decoder was presented by Ma et al. \cite{Ma11} where on the decoder side, dynamic voltage and frequency scaling is applied when the modeled processing complexity is small enough. A different approach using a high-level power model for a hardware-accelerated decoder is used by Xin Li et al. \cite{Li12}. In contrast, our approach uses energy values obtained from real measurements for a high number of different intraframe-coding tools such that more detailed and more accurate energy estimates are obtained. Furthermore, our work aims at optimizing the decoding energy at a constant objective visual quality.

The paper is organized as follows: Section \ref{sec:model} revisits the energy model that we use to obtain energy-aware encoding decisions. Afterwards, Section \ref{sec:DEDO} explains how this model is incorporated into the encoder and according to which criteria the encoding decisions are taken. Then, Section \ref{sec:eval} introduces our test setup and gives evaluation results for the proposed algorithm in a local playback and a streaming scenario. Section \ref{sec:concl} concludes the paper.

 
\section{Decoder Energy Model}
\label{sec:model}
The energy model that we use to describe the energy consumption of the decoder was presented in \cite{Herglotz14} and is applicable for various different decoding systems \cite{Herglotz15c}. It includes the processing energy of the CPU as well as the energy required by the random-access-memory (RAM). We adopt the accurate model to estimate the decoding energy for intra coded sequences that is calculated by 
\begin{align}
  \hat E_\mathrm{dec}  = \ & E_0 + e_\mathrm{slice}\cdot n_\mathrm{slice} \notag \\  
& +  \sum_\mathrm{\forall \mathrm{size}}{\left(\sum_{\forall \mathrm{mode}}{e_\mathrm{mode,size}}\cdot n_\mathrm{mode,size}\right)} \notag \\ 
& + \sum_{\forall \mathrm{size}} \left( \sum_{\forall \mathrm{comp}}  \!\!e_\mathrm{comp, size} \cdot n_\mathrm{comp, size}\right) \notag \\
& + e_\mathrm{coeff}\cdot n_\mathrm{coeff} + e_\mathrm{g1}\cdot n_\mathrm{g1} \notag  \\
& +  e_\mathrm{val}\cdot \sum_{\forall c\ne 0}{\log _2 \left| c\right|} + e_\mathrm{CSBF}\cdot n_\mathrm{CSBF} \notag \\
& + e_\mathrm{noMPM}\cdot n_\mathrm{noMPM} -  e_\mathrm{TSF}\cdot n_\mathrm{TSF}.  
\label{eq:intraModel}
\end{align}
In this model, the values of the variables $e$ (``specific energies'') represent the energy required for executing different functions during the decoding process. These functions can be executed multiple times. For example, the coefficient $e_\mathrm{coeff}$ can be interpreted as the energy required to decode a single non-zero residual coefficient, $e_\mathrm{g1}$ is the additional decoding energy if the coefficient's value is greater than one. Likewise, the variables $e_\mathrm{mode,size}$ describe the intra-prediction process on a certain block size that can range from $32\times 32$ to $4\times 4$. CSBF corresponds to the coded subblock flags, MPM to the most probable modes, and TSF to the transform skip flags. $e_\mathrm{comp, size}$, where comp represents the color components Y, U and V, represents the energy required for transformation of the residual coefficients. Further information about these specific energies can be found online \cite{denesto}. 

Counting how often these functions are executed during the decoding process of a single bit stream the so-called feature numbers $n$ can be determined. Multiplying these numbers $n$ with the specific energies $e$, we obtain an estimation for the complete required decoding energy. In \cite{Herglotz14} it was shown that estimation errors of less than $3\%$ with respect to the measured true energy consumption can be achieved. 

There are two important reasons for choosing this model: First, if specific energies are given, the decoding energy can be estimated without having to execute the decoding process or having the decoder at hand. The information about the bit stream features, which is inherently available in the encoder, is sufficient. Second, most of the specific energies can be assigned to encoding decisions. An example shall visualize this property: Consider the encoder needs to decide if residual coefficients for the luma component on a certain block size shall be coded or not. In the first option, a certain number $n_\mathrm{coeff}$ of quantized coefficients will be coded where some of them are greater than one ($n_\mathrm{g1}$). Furthermore, the inverse transformation of this block has to be performed once ($n_\mathrm{comp=luma,size}=1$). Multiplying the numbers with the corresponding specific energies and adding up the products we obtain an estimation on how much energy the decoder needs to process these tasks. For the second option, none of these tasks needs to be executed such that no additional energy is required. 

With the help of this approach, the encoder gets a third encoding criterion next to rate and distortion: the decoding energy that we exploit for Decoding-Energy-Rate-Distortion-Optimization (DERDO) in the next section.

\section{Decoding-Energy-Rate-Distortion Optimization}
\label{sec:DEDO}
In order to obtain savings in decoding energy we decided to modify the standard Rate-Distortion-Optimization (RDO) approach as presented in \cite{Sullivan98}. Therefore, we include the estimated decoding energy into the standard equation and obtain 
\begin{equation}
  \min J_\mathrm{DERD} = D + \lambda_\mathrm{R} R + \lambda_\mathrm{E} \hat E, 
  \label{eq:minJ_DERDO}
\end{equation}
where we adopt the parameters distortion $D$, rate $R$, and the corresponding Lagrange multiplier $\lambda_\mathrm{R}$ from the classic approach. In addition, we consider the estimated decoding energy $\hat E$ with a corresponding Lagrange multiplier $\lambda_\mathrm{E}$ that will be derived in the next subsection. $J_\mathrm{DERD}$ is the cost function to be minimized. During the encoding process, the decoding energy $\hat E$ is estimated at runtime using (\ref{eq:intraModel}).

Furthermore, we tested another approach neglecting the rate by calculating 
\begin{equation}
  \min J_\mathrm{DED} = D + \lambda_\mathrm{E} \hat E. 
  \label{eq:minJ_DEDO}
\end{equation}
The results to this approach will show the maximum potential energy savings. As the rate is not considered at all it can increase significantly. This approach is also used to obtain the $\lambda_\mathrm{E}$-QP relation as shown in the next subsection. 



 
\subsection{Lambda-QP Relation}
A major challenge in constructing a helpful optimization formula is finding a relation between the Lagrange multiplier $\lambda_\mathrm{E}$ and the quantization parameter (QP). For optimal coding results, the QP should be independent from the choice of $\lambda_\mathrm{E}$ such that an exhaustive search across different QP values would be necessary. To save encoding time, in standard RDO, it was shown that for a fixed $\lambda_\mathrm{R}$, a single QP is chosen in most cases \cite{Wiegand01, binli13}, such that a fixed equation relating $\lambda_\mathrm{R}$ to the QP is proposed as 
\begin{equation}
  \lambda_\mathrm{R} = c\cdot 2^{\frac{QP-12}{3}}, 
  \label{eq:lambda_QP_traditional}
\end{equation}
where $c=0.57$ has been determined experimentally and is used for the HM encoder implementation \cite{HM}. 

We adopt this approach for DERDO and experimentally determine the relation between the QP and $\lambda_\mathrm{E}$. As decoding energy model, we take specific energies for the HM implementation that will be introduced in Section \ref{sec:eval}. 

\begin{figure}
\centering
\psfrag{000}[c][c]{}
\psfrag{001}[c][c]{$10$}
\psfrag{002}[c][c]{$20$}
\psfrag{003}[c][c]{$30$}
\psfrag{004}[c][c]{$40$}
\psfrag{005}[c][c]{$50$}
\psfrag{006}[r][r]{$0$}
\psfrag{007}[r][r]{$0.5$}
\psfrag{008}[r][r]{$1$}
\psfrag{009}[l][c]{$5.84\cdot10^9$}
\psfrag{010}[l][c]{$1.84\cdot10^9$}
\psfrag{011}[l][c]{$5.79\cdot10^8$}
\psfrag{012}[l][c]{$1.82\cdot10^8$}
\psfrag{013}[l][c]{$5.75\cdot10^7$}
\psfrag{014}[l][c]{$1.81\cdot10^7$}
\psfrag{015}[l][c]{$5.70\cdot10^6$}
\psfrag{016}[l][c]{$1.80\cdot10^6$}
\psfrag{017}[l][c]{$5.65\cdot10^5$}
\psfrag{018}[b][c]{Relative occurence [\%]}
\psfrag{019}[t][c]{QP value}
\includegraphics[width=0.5\textwidth]{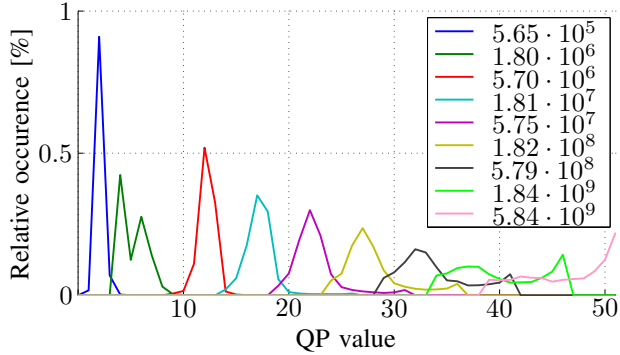}
\caption{Relative frequency of occurences of the QPs for the $\lambda_\mathrm{E}$-values shown in the legend. The tested input sequences are listed in Table \ref{tab:eval_vids} below. }
\label{fig:lambda_QP_relation}
\end{figure}

To get a first coarse approximation for $\lambda_\mathrm{E}$, we compared the decoding energy to the complete number of bits for several bit streams and found that the energy is usually lower by approximately six orders of magnitude (e.g., the sequence BasketballPass was encoded using $2.04$ MBits and requires $1.78$J for decoding). Hence, we multiplied the $\lambda$-values chosen in the reference software by $5\cdot10^6$ and tested eleven values for $\lambda_\mathrm{E}$ in the range of $[5.65\cdot10^5,5.84\cdot 10^9]$. To obtain the optimal QP, it was optimized on CTU-level in the corresponding range of $[5\pm 5, 45\pm5]$. The resulting relative occurences are depicted in Figure \ref{fig:lambda_QP_relation}. 

\begin{figure}
\centering
\psfrag{000}[c][b]{$0$}
\psfrag{001}[c][b]{$10$}
\psfrag{002}[c][b]{$20$}
\psfrag{003}[c][b]{$30$}
\psfrag{004}[c][b]{$40$}
\psfrag{005}[c][b]{$50$}
\psfrag{006}[r][r]{$10^4$}
\psfrag{007}[r][r]{$10^6$}
\psfrag{008}[r][r]{$10^8$}
\psfrag{009}[r][r]{$10^{10}$}
\psfrag{010}[r][r]{$10^{12}$}
\psfrag{011}[l][c]{\footnotesize{Eq. (\ref{eq:lambdaEfromQP})}}
\psfrag{012}[l][c]{\footnotesize{mostly chosen QPs}}
\psfrag{013}[b][c]{$\lambda_\mathrm{E}$}
\psfrag{014}[t][l]{QP value}
\includegraphics[width=0.4\textwidth]{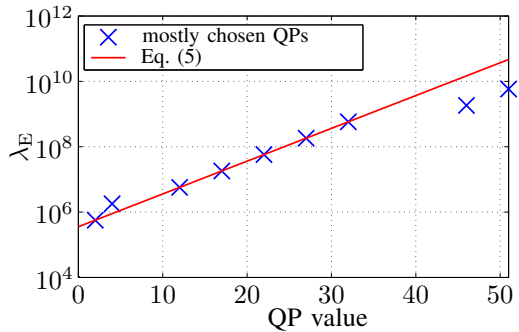}
\caption{Experimental relation between QP and $\lambda_\mathrm{E}$. The markers correspond to the maxima shown in Figure \ref{fig:lambda_QP_relation}, the curve is the proposed approximation.  }
\label{fig:lambda_QP_appr}
\end{figure}

We can observe that distinct peaks occur especially for lower values of $\lambda_\mathrm{E}$. For higher QPs the peaks shrink and even disappear. However, for simplification, we decided to stick to the traditional approach and found that 
\begin{equation}
\lambda_\mathrm{E} = c_\mathrm{E} \cdot 2^{\frac{QP-12}{3}} 
\label{eq:lambdaEfromQP}
\end{equation}
with $c_\mathrm{E} = 0.57\cdot 10^{7}$ is well suited to represent the relation. Figure \ref{fig:lambda_QP_appr} shows the experimental relation between the dominant QPs and the fixed $\lambda_\mathrm{E}$ as well as the curve described by (\ref{eq:lambdaEfromQP}). 
We can see that for the more common intermediate QPs, the experimental result is well approximated.

\section{Evaluation}
\label{sec:eval}
We prove the applicability of our approach by testing the HM-13.0 decoder solution \cite{HM} and the optimized decoder in the FFmpeg framework \cite{FFmpeg} on a Pandaboard \cite{Panda}. The Pandaboard is a development platform that is equipped with an OMAP-4430 System-on-Chip that is typical for portable devices like smartphones. To suppress impacts from background processes or peripheral hardware, we performed the decoding on runlevel 1 with disabled LEDs and disabled monitoring tasks. The measurement setup used to determine the processing energy of the pure decoding process is the same as in \cite{Herglotz15a}.

To estimate the decoding energy in the encoder, we used the accurate model with energy parameter values for the HM-decoder that can be found online \cite{denesto}. For encoding, we adapted the HM-14.0 encoder by incorporating the specific energies $e$ into the respective decision stages. To obtain the real resulting energy savings, the decoding energy for all resulting bit streams was newly measured. Hence, the results given in this section describe the real energy savings and not the savings estimated by the model which may be inaccurate. 

As input sequences, we chose a subset of the HEVC test set as shown in Table \ref{tab:eval_vids}. 
\begin{table}[t]
\renewcommand{\arraystretch}{1.3}
\caption{Properties of input sequences. The sequences are taken from the HEVC test set and were encoded beginning with the first frame with QPs ranging from $15$ to $45$ in steps of $5$.  }
\label{tab:eval_vids}
\vspace{-.3cm}
\begin{center}
\footnotesize{\begin{tabular}{l|c|c|r}
\hline
Sequence & Class & Resolution & No. frames \\
\hline
PeopleOnStreet (PoS) & A & $2560\times1600$ & 5 \\
Traffic (Tr) & A & $2560\times1600$ & 5 \\
Kimono (Ki) & B & $1920\times1080$ & 16 \\
RaceHorses (RHC)& C & $832\times480$ & 50 \\
BasketballPass (BP)& D & $416\times240$ & 50 \\
BlowingBubbles (BB)& D & $416\times240$ & 50 \\
BQSquare (BQ)& D & $416\times240$ & 50 \\
RaceHorses (RHD)& D & $416\times240$ & 50 \\
vidyo3 (vid)& E & $1280\times720$ & 50 \\
SlideEditing (SE)& F & $1280\times720$ & 50 \\
 \hline
\end{tabular}}
\end{center}
\vspace{-0.5cm}
\end{table}
The restricted number of frames was chosen to keep processing and measuring time at a reasonable level, where further tests indicated that coding more frames does not change the coding efficiency significantly. All sequences were encoded three times for optimal HM-decoding where all frames were coded as intra frames:  First, we used the standard RDO and second, the proposed DERDO approach as shown in (\ref{eq:minJ_DERDO}). In the third approach (DEDO), the rate was neglected as shown in (\ref{eq:minJ_DEDO}). 

We used the following three properties to compare the performance of both algorithms: First the YUV-PSNR that is calculated by 
\begin{equation}
  \mathrm{PSNR}_\mathrm{YUV} = \frac{1}{8}\left( 6\cdot \mathrm{PSNR}_\mathrm{Y} + \mathrm{PSNR}_\mathrm{U} + \mathrm{PSNR}_\mathrm{V} \right), 
  \label{eq:YUV_PSNR}
\end{equation}
second the bitrate $R$ in terms of bit stream file size, and third the measured complete decoding energy $E$ in Joule [J]. We evaluate the coding efficiency for two use cases: Local playback that focuses on the pure processing energy (Sec. \ref{secsec:locPlay}) and an online streaming application where transmission energies are considered (Sec. \ref{secsec:streaming}). 

\subsection{Local Playback}
\label{secsec:locPlay}

\begin{figure*}[ht]
\centering
\subfigure
{
\psfrag{000}[c][c]{$0$}
\psfrag{001}[c][c]{$20$}
\psfrag{002}[c][c]{$40$}
\psfrag{003}[r][r]{}
\psfrag{004}[r][r]{$40$}
\psfrag{005}[r][r]{$50$}
\psfrag{006}[r][c]{}
\psfrag{007}[r][c]{$0$}
\psfrag{008}[r][c]{$1$}
\psfrag{009}[r][c]{$2$}
\psfrag{010}[r][r]{$30$}
\psfrag{011}[r][r]{$40$}
\psfrag{012}[r][r]{$50$}
\psfrag{014}[l][c]{\footnotesize{DERDO}}
\psfrag{015}[l][c]{\footnotesize{DEDO}}
\psfrag{016}[l][c]{\footnotesize{RDO}}
\psfrag{018}[b][t]{$\mathrm{PSNR}_\mathrm{YUV}$}
\psfrag{017}[t][c]{Bitrate increase [\%]}
\psfrag{013}[t][l]{Bitrate [MBits]}
\psfrag{019}[t][l]{}
\includegraphics[width=0.48\textwidth]{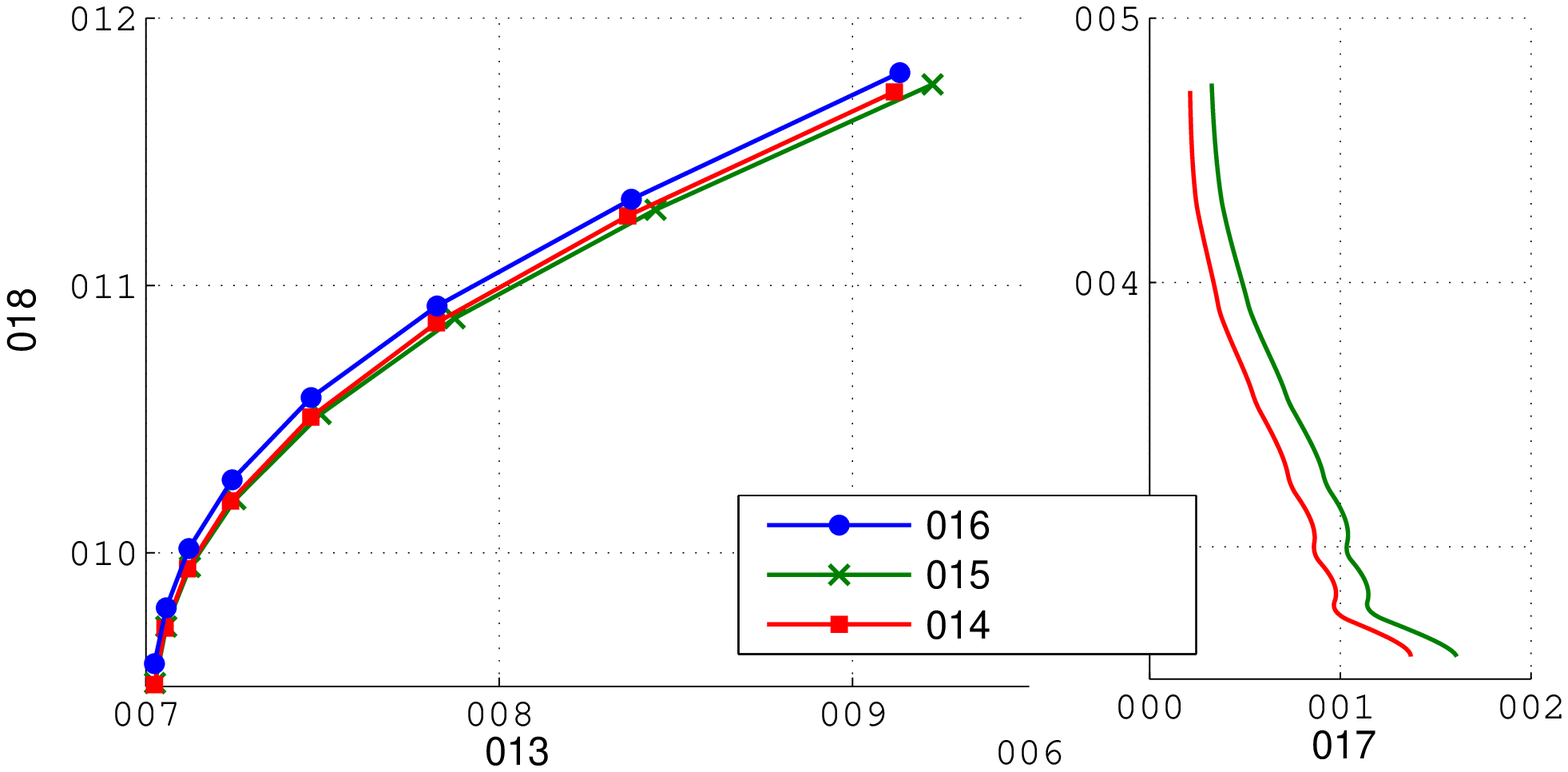} 
\label{fig:RD_bubbles}
}
\subfigure
{
\psfrag{000}[c][c]{$-30$}
\psfrag{001}[c][c]{$-15$}
\psfrag{002}[c][c]{$0$}
\psfrag{003}[r][r]{}
\psfrag{004}[r][r]{$40$}
\psfrag{005}[r][r]{$50$}
\psfrag{006}[r][c]{$0$}
\psfrag{007}[r][c]{$1$}
\psfrag{008}[r][c]{$2$}
\psfrag{009}[r][c]{$3$}
\psfrag{010}[r][c]{$4$}
\psfrag{011}[r][c]{$5$}
\psfrag{012}[r][r]{$30$}
\psfrag{013}[r][r]{$40$}
\psfrag{014}[r][r]{$50$}
\psfrag{015}[r][r]{$50$}
\psfrag{015}[l][c]{\footnotesize{DERDO}}
\psfrag{016}[l][c]{\footnotesize{DEDO}}
\psfrag{017}[l][c]{\footnotesize{RDO}}
\psfrag{019}[b][t]{$\mathrm{PSNR}_\mathrm{YUV}$}
\psfrag{018}[t][t]{Energy saving [\%]}
\psfrag{020}[t][t]{Decoding energy [J]}
\includegraphics[width=0.48\textwidth]{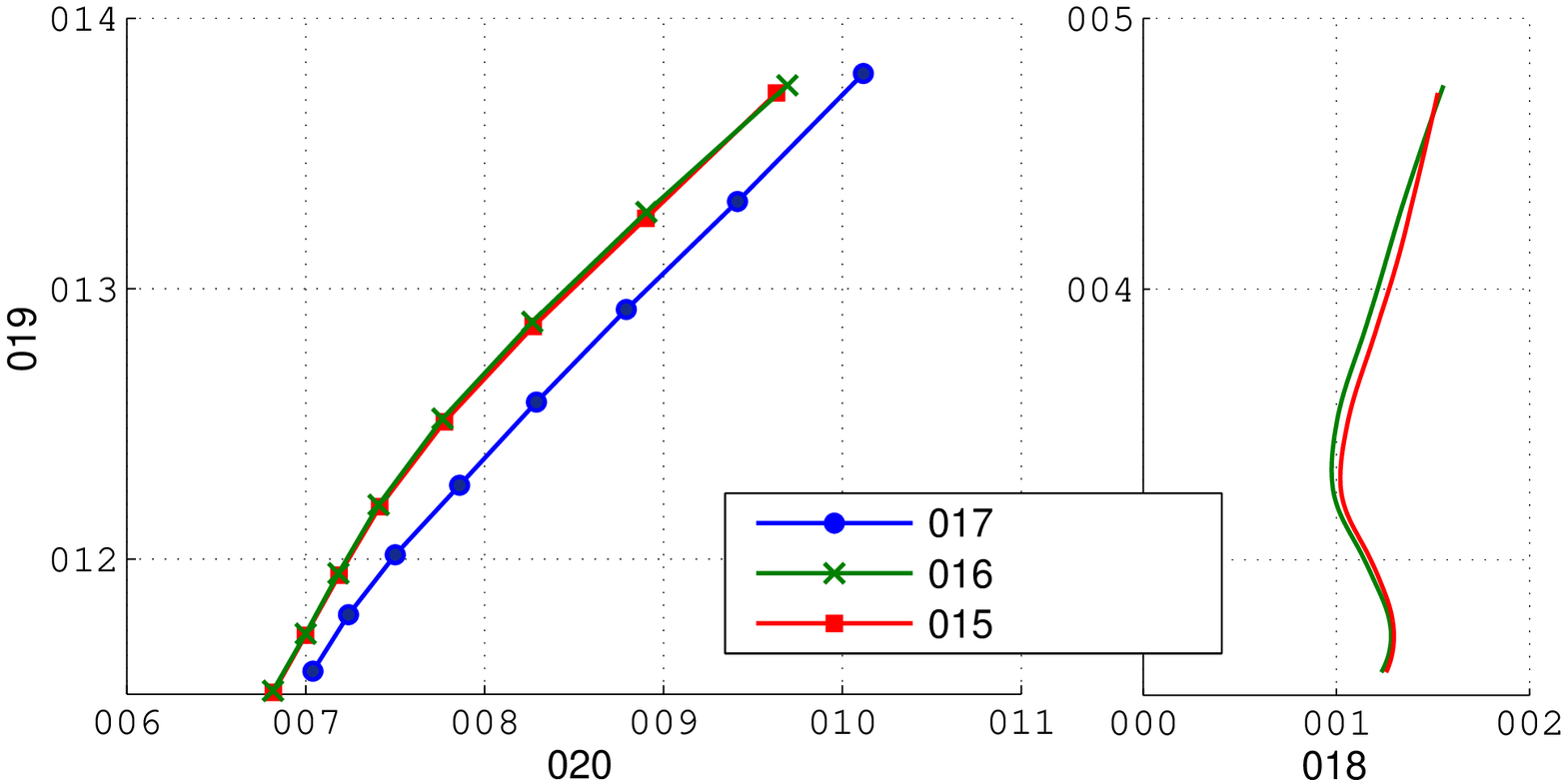}
\label{fig:DED_bubbles}}
\caption{Rate-Distortion (a) and Decoding-energy-distortion (b) performance of the standard RDO (blue lines) and the proposed DEDO (green lines) and DERDO (red lines) approach for the BlowingBubbles sequence decoded with the HM-software. The left diagrams in each subfigure relate distortion with rate and energy, respectively, where the markers depict the QPs ranging from $15$ to $50$ in steps of $5$. The right diagrams in each subfigure show the relative bitrate increases and energy savings in percentage in comparison to the standard RDO. }
\vspace{-.4cm}
\end{figure*}

The results given in this section describe the energy savings in a local playback scenario. Figures \ref{fig:RD_bubbles} and \ref{fig:DED_bubbles} compare the RD-performance and the energy saving performance of the proposed approaches for the BlowingBubbles sequence using the HM-decoder. 
In Figure \ref{fig:RD_bubbles} we can see that as expected, using the new minimization functions results in losses in RD-performance. For the complete QP range, the RD-curves from DEDO and DERDO lie below the curve from RDO. To visualize how many more bits have to be spent to achieve the same visual quality, the right diagram gives the relative amount of extra bits needed for the proposed approach. We can see that especially for high QPs (low image qualities) more bits are required. 

In contrast, regarding Fig. \ref{fig:DED_bubbles} we can see that a significant amount of decoding energy can be saved. The right diagram indicates possible savings of up to $15\%$. The curves of the other sequences show a similar behavior, though we observed that they can be shifted towards higher or lower rates and energies depending on their content. 
Furthermore, we can see that in terms of energy savings, the DEDO approach only performs slightly better than the DERDO approach. As in contrast DERDO shows a significantly lower bitrate increase, this approach seems to be more appropriate for practical use. 

To summarize the results, we calculated the Bj{\o}ntegaard-Delta bitrate (BDBR) as proposed in \cite{BD_PSNR} for all sequences and both decoders. Furthermore, to give mean values for the energy savings, we use the same approach but replace the bitrate by the consumed energy and calculated a BD-decoding energy (BDDE). The resulting values for the HM decoder and the FFmpeg decoder are summarized in Table \ref{tab:BDvals}.

\begin{table}[t]
\renewcommand{\arraystretch}{1.3}
\caption{Average bitrate increases (BDBR) and decoding energy savings (BDDE) as calculated by the Bj{\o}ntegaard-Delta approach (QPs $15$, $25$, $35$, and $45$) for the HM and the FFmpeg decoder. }
\label{tab:BDvals}
\vspace{-0.3cm}
\begin{center}
\footnotesize{\begin{tabular}{l||r|r|r||r|r|r}
\hline
 & \multicolumn{3}{|c||}{DEDO} & \multicolumn{3}{|c}{DERDO} \\
\hline
Seq. & BDBR & \multicolumn{2}{|c||}{BDDE} & BDBR & \multicolumn{2}{|c}{BDDE} \\
\hline
 & & \multicolumn{1}{c|}{HM} & \multicolumn{1}{c||}{FFmpeg} & & \multicolumn{1}{c|}{HM} & \multicolumn{1}{c}{FFmpeg} \\
 \hline
PoS 	& $20.6\%$ & $17.4\%$ & $11.5 \%$ & $16.4\%$ & $16.7\%$ & $10.3 \%$\\
Tr 		& $18.9\%$ & $16.9\%$ & $14.9 \%$ & $15.3\%$ & $16.7\%$ & $14.2 \%$\\
Ki 		& $25.6\%$ & $10.6\%$ & $21.4\%$ & $21.3\%$ & $10.1\%$ & $19.2\%$\\
RHC		& $15.9\%$ & $15.0\%$ & $12.5 \%$ & $12.5\%$ & $14.5\%$ & $11.0 \%$\\
BP 		& $22.8\%$ & $13.5\%$ & $12.8 \%$ & $16.9\%$ & $12.9\%$ & $11.7 \%$\\
BB 		& $16.3\%$ & $15.8\%$ & $11.9 \%$ & $12.4\%$ & $15.1\%$ & $10.8 \%$\\
BQ 		& $13.9\%$ & $9.9\%$  & $7.6 \%$ & $9.9\%$  & $8.9\%$ & $5.9 \%$\\
RHD		& $17.6\%$ & $14.8\%$ & $10.3 \%$ & $13.0\%$ & $13.8\%$ & $9.4 \%$\\
vid 	& $28.7\%$ & $14.9\%$ & $17.4 \%$ & $23.1\%$ & $14.2\%$ & $16.2 \%$\\
SE 		& $10.4\%$ & $6.3\%$  & $5.2 \%$ & $6.7\%$  &  $5.7\%$ & $5.5 \%$\\
 \hline
\end{tabular}}
\end{center}
\vspace{-.6cm}
\end{table}

We can see that most energy can be saved for high resolution sequences. Furthermore, optimizing the bit stream for HM-decoding is also beneficial for a different decoding solution (FFmpeg), though savings are slightly lower (about $2\%$ in average). Interestingly, in some cases (sequence Ki and vid) savings are higher for the FFmpeg solution, which can be explained by inaccuracies in the energy model and the estimated specific energies. Summing up, when using the DERDO approach, accepting bitrate increases of $6\%$ to $24\%$, decoding energy can be saved in the range of $5\%$ to $17\%$. 

\subsection{Online Streaming}
\label{secsec:streaming}
In this subsection, in order to show how streaming of the video bit stream 
affects the overall energy consumption, we consider the transmission energy in a WiFi-streaming scenario. Therefore, we take the energy model proposed in \cite{Sun14}. In this model, a per-bit transmission energy $E_\mathrm{b}$ in nJ per bit can be estimated depending on the throughput $Th$ in megabit per second as 
\begin{equation}
\hat E_\mathrm{b} = a\cdot Th^{-1}+b, 
\end{equation}
where we take the values for parameter $a=305.3$ and $b=13.1$ for a universal transmission model (independent from protocol and packet size, values valid for a Google Nexus smartphone). Analog to our approach estimating the processing energy, we do not consider the idle transmission energy required for sustaining the network connection. 

To estimate the energy required in the streaming scenario, we calculate the throughput $Th$ of each test sequence (which is the product of the bit stream file size $R$ with the frame rate divided by the number of frames), derive the corresponding per-bit energy $\hat E_\mathrm{b}$ ($23-8000$nJ per bit), and add the estimated transmission energy ($\hat E_\mathrm{tr}=R\cdot \hat E_\mathrm{b})$ to the measured decoding energy. The resulting energy values are used to calculate the Bj{\o}ntegaard-Delta values shown in Table \ref{tab:BDvals_transmission}. 
\begin{table}[t]
\renewcommand{\arraystretch}{1.3}
\caption{Average bitrate increases and decoding energy savings as calculated by the Bj{\o}ntegaard-Delta approach (QPs $15$, $25$, $35$, and $45$) in an online streaming application (the BDBR-values are the same as in Table \ref{tab:BDvals}). }
\label{tab:BDvals_transmission}
\vspace{-0.3cm}
\begin{center}
\footnotesize{\begin{tabular}{l||r|r|r||r|r|r}
\hline
 & \multicolumn{3}{|c||}{DEDO} & \multicolumn{3}{|c}{DERDO} \\
\hline
Seq. & BDBR & \multicolumn{2}{|c||}{BDDE} & BDBR & \multicolumn{2}{|c}{BDDE} \\
\hline
 & & \multicolumn{1}{c|}{HM} & \multicolumn{1}{c||}{FFmpeg} & & \multicolumn{1}{c|}{HM} & \multicolumn{1}{c}{FFmpeg} \\
 \hline
PoS 	& $20.6\%$ & $17.2\%$ & $11.1 \%$ & $16.4\%$ & $15.6\%$ & $10.0 \%$\\
Tr 		& $18.9\%$ & $16.5\%$ & $14.5 \%$ & $15.3\%$ & $15.6\%$ & $13.8 \%$\\
Ki 		& $25.6\%$ & $9.2\%$ & $19.9\%$ & $21.3\%$ & $7.9\%$ & $17.9\%$\\
RHC		& $15.9\%$ & $14.0\%$ & $10.1 \%$ & $12.5\%$ & $12.9\%$ & $8.9 \%$\\
BP 		& $22.8\%$ & $11.4\%$ & $8.0 \%$ & $16.9\%$ & $10.2\%$ & $7.3 \%$\\
BB 		& $16.3\%$ & $14.2\%$ & $8.3 \%$ & $12.4\%$ & $12.9\%$ & $7.6 \%$\\
BQ 		& $13.9\%$ & $9.4\%$  & $5.5 \%$ & $9.9\%$  & $7.8\%$ & $4.3 \%$\\
RHD		& $17.6\%$ & $12.2\%$ & $6.1 \%$ & $13.0\%$ & $10.8\%$ & $5.6 \%$\\
vid 	& $28.7\%$ & $13.7\%$ & $16.1 \%$ & $23.1\%$ & $12.0\%$ & $15.0 \%$\\
SE 		& $10.4\%$ & $6.4\%$  & $4.6 \%$ & $6.7\%$  &  $5.5\%$ & $4.9 \%$\\
 \hline
\end{tabular}}
\end{center}
\vspace{-.6cm}
\end{table}
We can see that energy savings are about $0.3-3\%$ lower, where especially low resolution content is affected. For high resolution sequences, due to a much lower per-bit energy resulting from higher throughput, the transmission energy plays a minor role.

\section{Conclusion}
\label{sec:concl}
In this paper we showed that it is possible to encode decoding-energy saving bit streams when the energetic properties of the decoding system are known to the encoder. Decoding energy can be saved in the range of $5\%$ to $17\%$ at a constant objective visual quality when accepting a compression efficiency loss of $6\%$ to $24\%$  in terms of rate-distortion performance. 
Online streaming only slightly affects the energy savings. Further work will extend this concept to interframe coding, test it on other decoding systems, and consider the transmission energy directly in the optimization process.

\section*{Acknowledgment}

This work was financed by the Research Training Group 1773 ``Heterogeneous Image Systems'', funded by the German Research Foundation (DFG).
\bibliographystyle{IEEEtran}

\bibliography{IEEEabrv,literatureNeu}

\end{document}